\documentclass[aps,pra,10pt,twocolumn]{revtex4}
\usepackage{graphicx,graphics,wrapfig,rotating}         
\usepackage{bm,fancybox}
\usepackage{amsmath,amssymb}                          
\usepackage{times,euscript,oldgerm}              %
\usepackage[english]{babel}                      %
\usepackage{psfrag}
\usepackage{color}
\usepackage{epsfig}
\usepackage{dcolumn}

\usepackage[draft]{todonotes}

\newcommand{\unit}[1]{\ensuremath{\,\mathrm{#1}}}

\newcommand{\nm}{\unit{nm}}
\newcommand{\um}{\unit{\mu m}}
\newcommand{\ps}{\unit{ps}}
\newcommand{\s}{\unit{s}}
\newcommand{\V}{\unit{V}}

\newcommand{\dBm}{\unit{dBm}}
\newcommand{\dB}{\unit{dB}}

\begin{document}

\title{The Path to Increasing the Coincidence Efficiency of \\ Integrated Photon Sources}
\author{C. C. Tison$^1$$^,$$^2$$^,$$^3$, J. A. Steidle$^4$, M. L. Fanto$^3$$^,$$^4$, Z. Wang$^4$, N. A. Mogent$^4$, A. Rizzo$^5$,  S. F. Preble$^4$, P. M. Alsing$^3$}

\affiliation{$^1$ Florida Atlantic University, Boca Raton, Fl, 33431, USA  \\
             $^2$ Quanterion Solutions Incorporated, Utica, NY, 13502, USA \\
             $^3$ Air Force Research Laboratory, Information Directorate, Rome, NY, 13441, USA\\
             $^4$Rochester Institute of Technology, Rochester, NY, 14623, USA\\
             $^5$Haverford College, Haverford, PA, 19041, USA\\
	     $*$ C.C.T. and J.A.S. equally contributed to this research.\\}
\date{\today}

\begin{abstract}

Silicon ring resonators are used as photon pair sources by taking advantage of silicon's large third order nonlinearity with a process known as spontaneous four wave mixing. These sources are capable of producing pairs of indistinguishable photons but typically suffer from an effective $50\%$ loss. By slightly decoupling the input waveguide from the ring, the drop port coincidence ratio can be significantly increased with the trade-off being that the pump is less efficiently coupled into the ring. Ring resonators with this design have been demonstrated having coincidence ratios of $\sim 96\%$ but requiring a factor of $\sim 10$ increase in the pump power. Through the modification of the coupling design that relies on additional spectral dependence, it is possible to achieve similar coincidence ratios without the increased pumping requirement. This can be achieved by coupling the input waveguide to the ring multiple times, thus creating a Mach-Zehnder interferometer. This coupler design can be used on both sides of the ring resonator so that resonances supported by one of the couplers are suppressed by the other. This is the ideal configuration for a photon-pair source as it can only support the pump photons at the input side while only allowing the generated photons to leave through the output side. Recently, this device has been realized with preliminary results exhibiting the desired spectral dependence and with a coincidence ratio as high as $\sim 97\%$ while allowing the pump to be nearly critically coupled to the ring. The demonstrated near unity coincidence ratio infers a near maximal heralding efficiency from the fabricated device. This device has the potential to greatly improve the scalability and performance of quantum computing and communication systems.
\end{abstract}

\maketitle 

\section{Introduction}

Silicon photonics is proving to be a very promising platform for quantum information processing. Microring resonators are becoming a key component of such systems as they have been shown to be effective as photon-pair sources by means of spontaneous four wave mixing (SFWM)\cite{Dinu2003,Wakabayashi2015,Azzini2012,Engin2013,Clemmen2009,Steidle2015,Silverstone2014,
Silverstone2015,Harris2014,Preble2015}.
Often, it is desirable to have precisely one photon. While SFWM sources generate pairs of photons, single photons can be achieved through heralding. Heralding is a technique in which the detection of a single photon from a pair is used to determine the existence of the other. One of the fundamental issues with ring resonators is their inherent $50\%$ loss when critically coupled, regardless of operation in a single bus or double bus configuration\cite{Hach2014,Vernon2015,Vernon2016}. For single bus resonators, half of the generated photons are lost to scattering within the cavity. Double bus resonators are slightly different as the photons are free to leave the ring through either port - resulting in an effective loss of $50\%$. All of this assumes that the ring resonator is critically coupled to straight waveguides. By straying from this standard coupling scheme, it is possible to dramatically decrease the effective loss of the source. Presented here, are experimental results for two different coupling schemes that achieve greatly reduced photon-pair loss when compared to the critically coupled ring resonator.

\section{Asymmetric Gap Microring Resonator}
\subsection{Dual Bus Microring Resonator Theory}
\label{sec:Theory}

\begin{figure}[b]
\centering
\includegraphics[width=8cm]{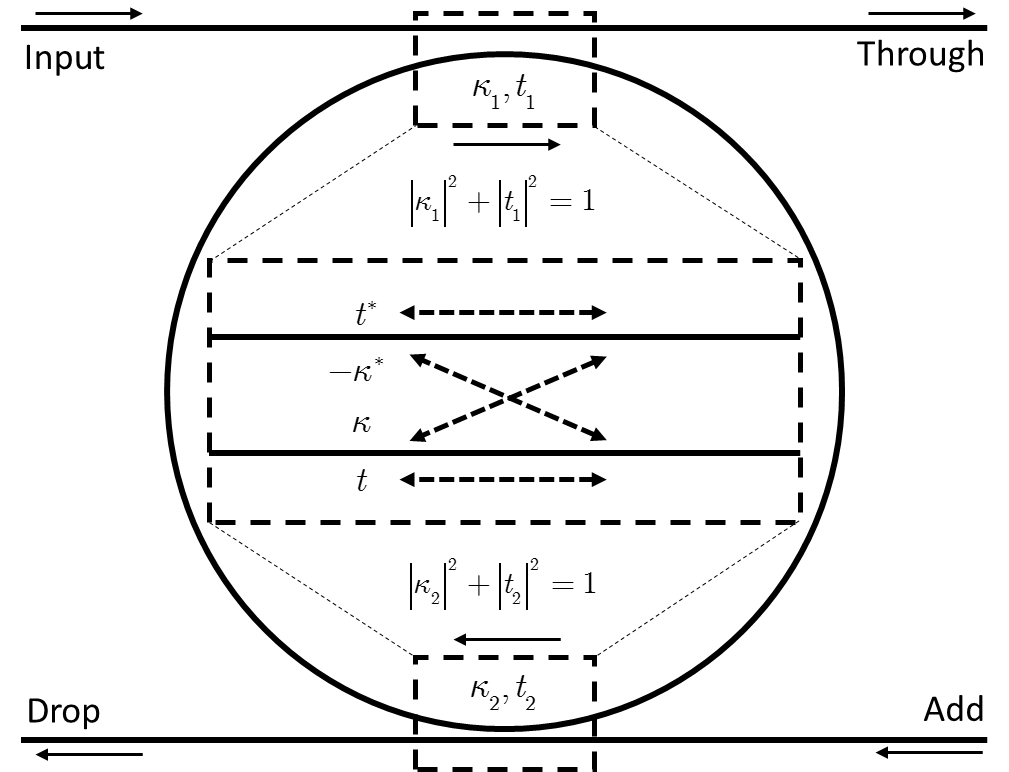}
\caption{Diagram of an asymmetrically coupled double bus ring resonator with an enlarged schematic of the evanescent couplers.}
\label{fig:ring}
\end{figure}

\begin{figure*}[ht]
\centering
\includegraphics[width=18cm]{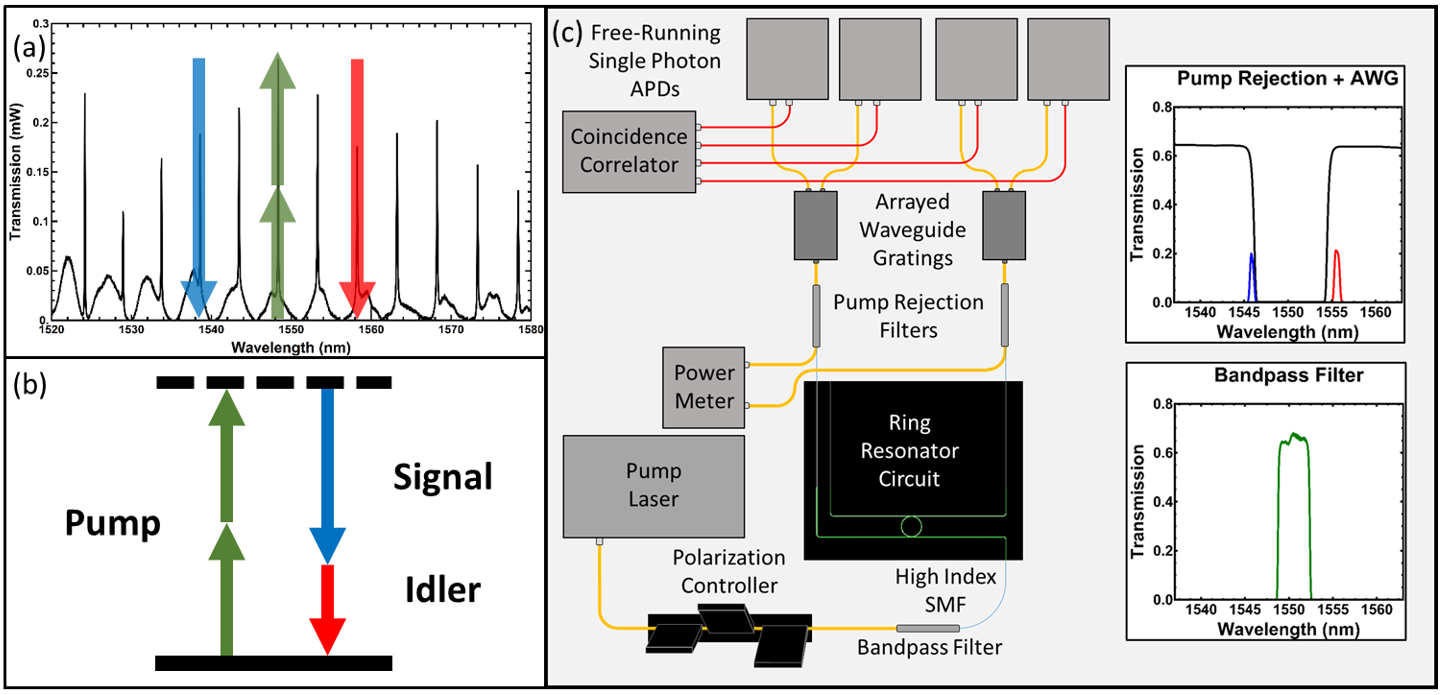}
\caption{(a) Transmission spectrum of the ring resonator source from the drop port of the device with a $150 \nm$ input gap. The arrows indicate the locations of the pump, signal, and idler photons. (b) Energy conservation schematic for the non-degenerate SFWM configuration that was used. (c) Schematic of the experimental setup along with plots of the filter transmissions.}
\label{fig:setup}
\end{figure*}

One method for increasing the coincidence efficiency of a microring resonator is to slightly decouple the input waveguide from the ring. A diagram of an asymmetrically coupled double bus ring resonator is shown in Fig. \ref{fig:ring}. Assuming the ring itself to be lossless, the probability of a photon generated inside the ring to couple out the drop port is the ratio between the input and output couplers given as

\begin{equation}\label{eq:pdrop}
p_\text{drop}=\frac{\left|\kappa_2\right|^2}{\left|\kappa_1\right|^2+\left|\kappa_2\right|^2}=\frac{1-\left|t_2\right|^2}{2-\left|t_1\right|^2-\left|t_2\right|^2}.
\end{equation}

Therefore, photons generated within a symmetric double bus ring resonator (i.e. $t_1 = t_2$) will exit both ports with equal probability. We can quantify the effect of loss through undesired ports of the ring through the relation



\begin{equation}\label{eq:pcoin}
\eta_\text{coinc}=\frac{{p_\text{drop}}^2}{\left( p_\text{thru}+p_\text{drop}\right)^2}=\left(\frac{1-\left|t_2\right|^2}{2-\left|t_1\right|^2-\left|t_2\right|^2}\right)^2.
\end{equation}

Evident by the above equation describing the coincidence ratio, an increase in the gap between the input waveguide and the ring (an increase in $\left|t_1\right|^2$) raises the percentage of drop port coincidences from $25\%$ (symmetric coupling) to near unity. Additionally, heralding efficiency will be proportional to this trend. The drawback of decreasing through port coupling is that the pump will under couple into the ring, resulting in an increase in required pump power to achieve the same level of ring excitation. This is an acceptable trade off as it is much easier to increase the pump power than it is to increase the brightness of the microring source. Recently, Vernon et. al worked through the complete theory for a single bus ring, reaching a similar conclusion \cite{Vernon2016}.

\subsection{Asymmetric Gap Microring Experimental Procedure}
\label{sec:Exp}

In this experiment, illustrated in Figure \ref{fig:setup},  a tunable continuous wave pump laser was set to the resonant wavelength (of the microring cavity) closest to $1550\nm$. The linearly polarized pump passed through a series of bandpass filters to minimize the excess laser noise injected into the system. Coupling efficiency was increased by utilizing an inverse taper on the silicon waveguide and fusion splicing a short section of high index fiber (Nufern UHNA-7) to the optical fiber (SMF-28) which was buttcoupled to the chip. Once on the chip, the pump coupled into the ring resonator ($R = 18.5 \um$, $W = 500\nm$, and $H = 220 \nm$) where non-degenerate photon pairs were produced at resonances spaced symmetrically about the pumped resonance. The generated photons then left the resonator through either port before coupling back into fiber. A series of notch filters (pump rejection) were used (for both outputs of the chip) to remove any remaining pump photons. A subsequent set of arrayed waveguide gratings (AWGs) spectrally separated the non-degenerate photon pairs into different fibers that each led to a free running single photon detector (ID Quantique ID230). Time correlation measurements were then collected with a time to digital converter (ID Quantique ID800) receiving signals from each of the four detectors.

Coincidence measurements were made for multiple devices with varying gaps (input gaps ranging from $150\nm$ to $350\nm$ with a constant drop port gap of $150\nm$). For each of these, coincidences were integrated for $900\s$ with a timing resolution of $81\ps$ for pump powers ranging from $-5\dBm$ to $5\dBm$. The variation in pump power enabled determination of the difference in the required pump power between devices to achieve the same level of ring excitation.

\subsection{Asymmetric Gap Microring Coincidence Efficiency Measurements}
\label{sec:Results}

\begin{figure}[t]
\centering
\includegraphics[width=8cm]{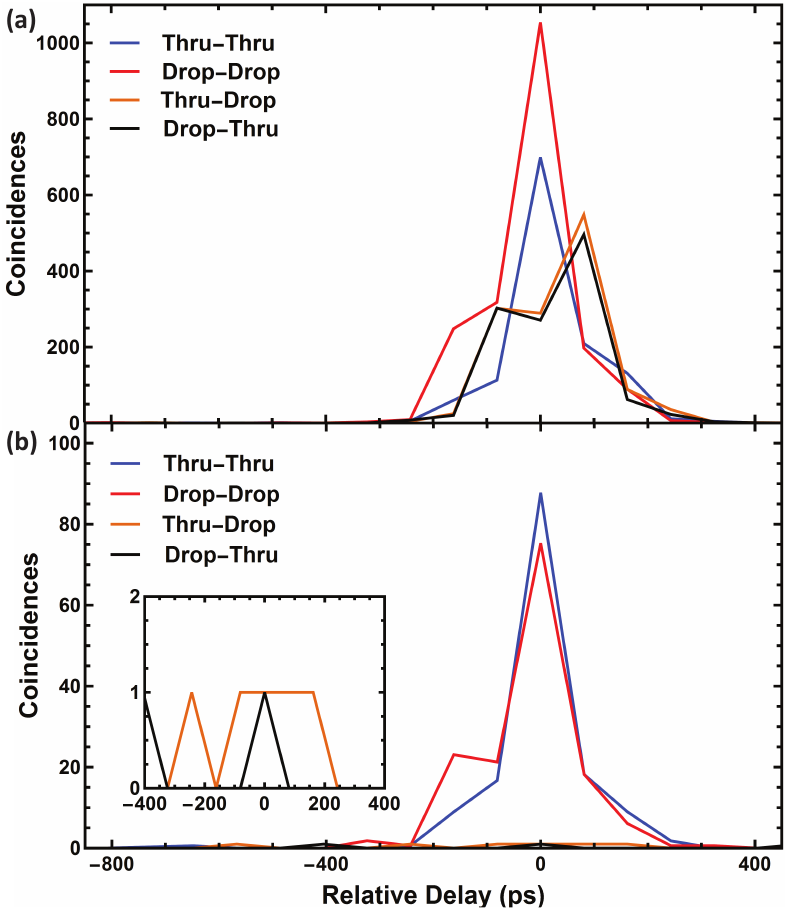}
\caption{(a) Coincidence plots for a device with a $150\nm$ input gap and a pump power of $5\dBm$. (b) Coincidence plots for a device with a $350\nm$ input gap and a pump power of $5\dBm$. The inset shows a zoomed in view of the coincidence plots for the cases where the photon pairs are split. }
\label{fig:gap_coincidences}
\end{figure}

As the waveguide-resonator gap on the input side of the ring was incrementally increased, the measured cross coincidences (through-drop) rapidly decreased, matching the expectation that the photons would only exit out the drop port (Fig. \ref{fig:gap_coincidences}). However, the coincidences from the through port remained high. The cause of this was determined to be the generation of photons in the input waveguide leading up to the ring resonator, which is a broadband four wave mixing process. To prove this, the pump laser was tuned to be off-resonance (eliminating the possibility of photon generation within the ring) and coincidences were collected. It was immediately obvious that the through port coincidence counts between the two cases (on-resonance and off-resonance) were very comparable to each other while the drop and split coincidences were approximately equal to zero in the off-resonance case. Therefore, to remove the effect of photon generation from the input waveguide, it was necessary to determine the coincidence ratio of the source using the drop port coincidences ($C_\text{drop,drop}$) and the split coincidences ($C_\text{thru,drop}$ and $C_\text{drop,thru}$) only using the following equation:

\begin{equation}
\eta_\text{coinc}=\frac{{C_\text{drop,drop}}^2}{\left(C_\text{drop,drop}+\frac{C_\text{thru,drop}+C_\text{drop,thru}}{2}\right)^2}.
\end{equation}

In order to get accurate values for the coincidence ratio, corrections needed to be made to the raw data from each device. First, the variation in fiber-chip coupling efficiency between the two outputs of the silicon chip was addressed by analyzing the spectrum of the ring resonator from both the through and drop ports. The other source of error to correct for was the variation in the loss between the four different optical paths leading from the chip to the detectors. This was addressed by collecting additional sets of data for each device after swapping the paths that the through and drop port photons took after they exited the chip.

\begin{figure}[b]
\centering
\includegraphics[width=8cm]{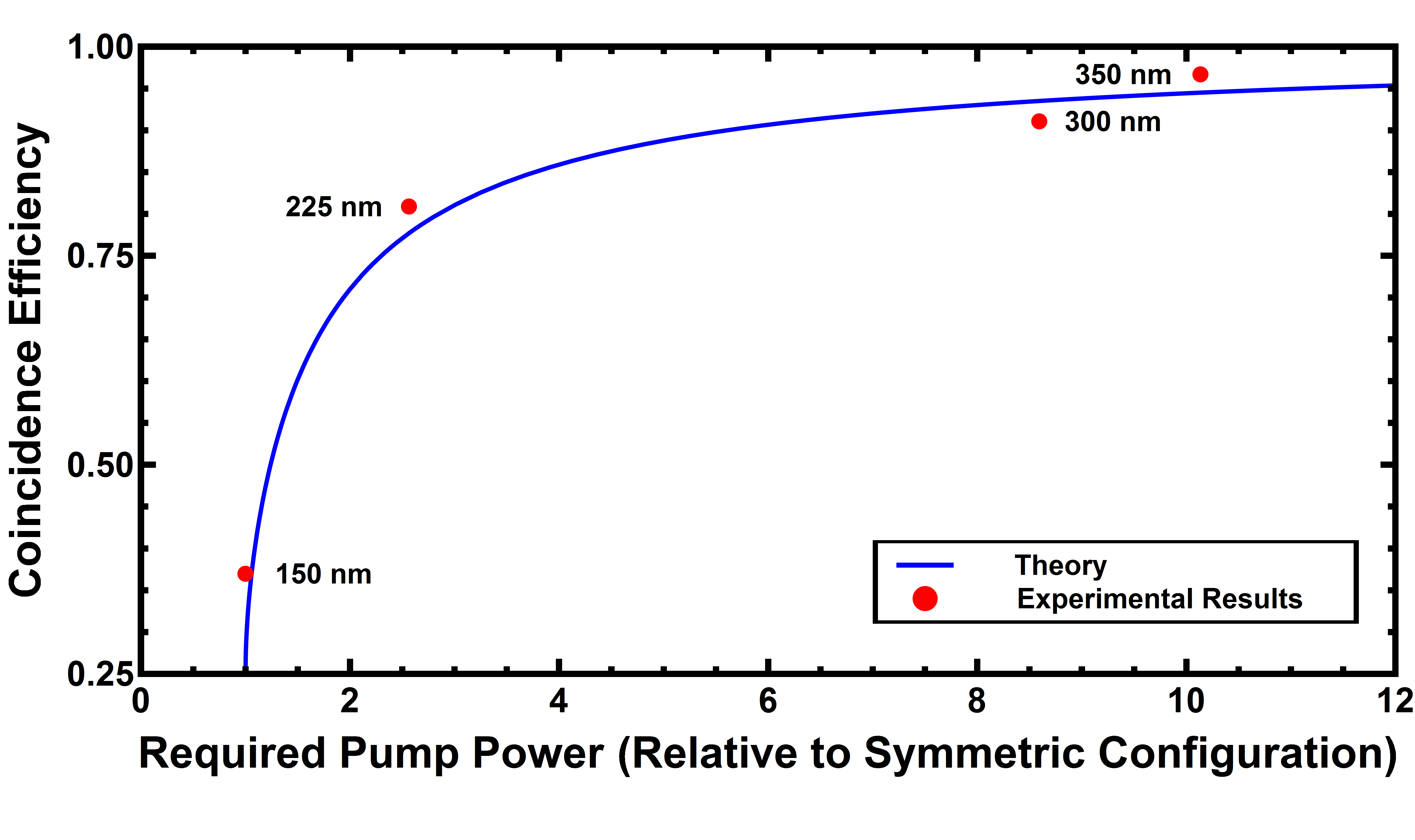}
\caption{Comparison between the experimental results and the theory. The size of the gap between the input waveguide and the ring is labeled for each data point.}
\label{fig:gap_results}
\end{figure}

The resulting coincidence plots (shown in Fig. \ref{fig:gap_coincidences}) clearly show that the drop port coincidences remain significant (the large drop in total coincidences is due to the decreased coupling of the pump into the ring) while the split coincidences (thru-drop and drop-thru) decreased to approximately zero between a device with symmetric $150\nm$ gaps and one with a $350\nm$ input gap. Comparing the ring spectra of the various devices allowed the determination of how efficiently the pump was coupling into the ring relative to the critically coupled device. This allowed proper comparison of all of the devices (Fig. \ref{fig:gap_results}). The symmetric device was found to have a relative coincidence ratio (\ref{eq:pcoin}) of $\eta_\text{coinc}=0.370$. While this is significantly higher than the theoretical value of $0.25$, a very small imperfection in the device fabrication can easily result in that ring-gap variation as it is at the most sensitive point on the theory curve. Increasing the input gap to $225\nm$ resulted in a device with a largely improved coincidence ratio of $0.809$ ($2.19\, x$ improvement) and requiring a factor of $2.56$ increased pump power. Further improvements were made with input gaps of $300\nm$ and $350\nm$ having coincidence ratios of $0.911$ ($2.46\, x$ improvement) and $0.967$ ($2.61\, x$ improvement) respectively while requiring increased pumping by factors of $8.60$ and $10.1$ respectively. As can be seen in Figure \ref{fig:gap_results}, these results are in agreement with the theory. While the increase in required pump power is an acceptable trade-off for the improved coincidence ratio, the additional pumping requirement can be removed by taking a different approach to the problem.

\section {Dual Mach-Zehnder Microring Resonator}
\subsection{Dual Mach-Zehnder Microring Theory}
\label{sec:DMZR_Theory}

In 1995, Barbarossa \emph{et~al.} found that resonant wavelengths of a microring cavity could theoretically be suppressed by coupling the input waveguide to the ring at two points \cite{Barbarossa1995}. This design essentially makes a Mach-Zehnder interferometer (MZI) out of the input waveguide and the ring and has since been demonstrated experimentally \cite{Chen2007,Popovic2008,Zeng2015}. To understand how such a device will operate, it is useful to first think of the individual components.

\begin{figure}[b]
\centering
\includegraphics[width=8cm]{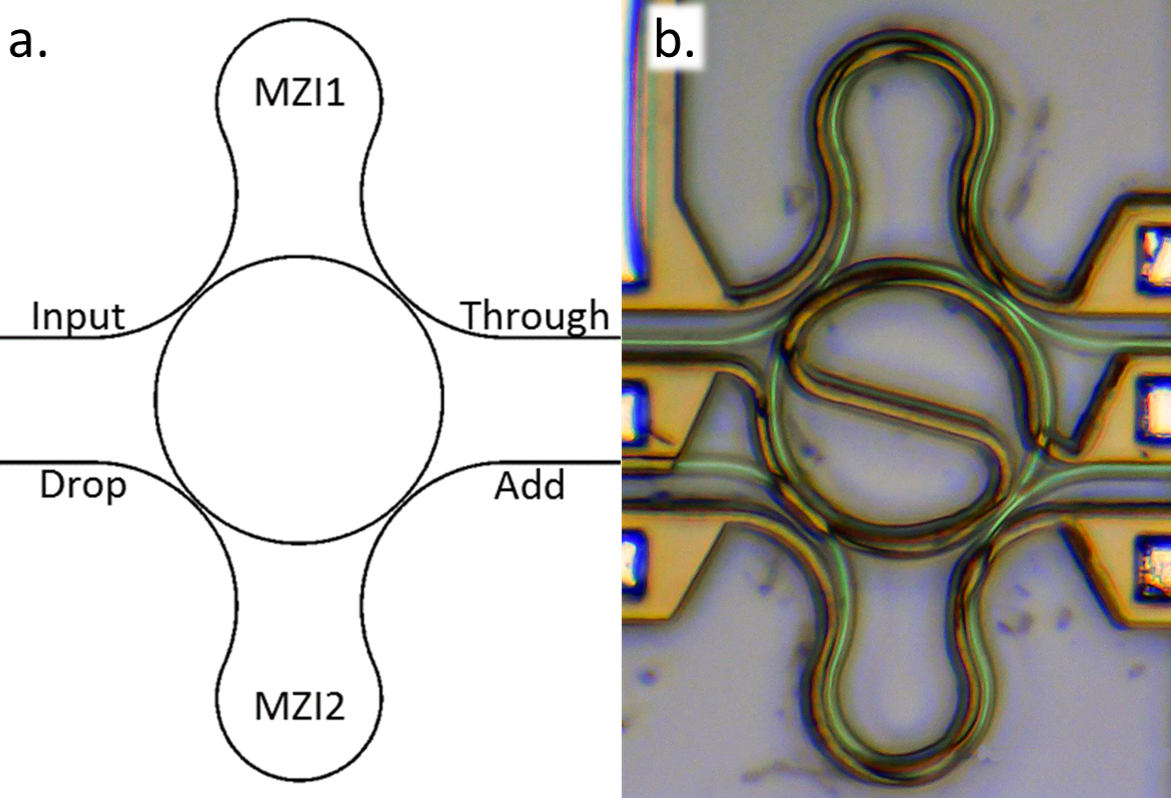}
\caption{(a) Schematic and (b) optical microscope image of the fabricated microring source with Mach-Zehnder interferometer couplers.}
\label{fig:dmzr_schematic}
\end{figure}

Being a cavity, the ring will only support specific wavelengths of light (where the resonance condition is satisfied) separated by the free spectral range (FSR). The spectrum of an unbalanced MZI is sinusoidal with the difference in optical path length between the two paths determining where in the spectrum the constructive and destructive interference will occur. For both the ring and the MZI, this is known as phase-matching. For the case of the ring this is phase-matching between consecutive round-trips while in the MZI it is phase-matching between the two different paths. The points of constructive interference in the spectra of these devices can be tuned by adjusting the relative phase between the different paths. In a fabricated device this can be accomplished by heaters or electro-optic phase shifters. The combination of these two elements results in a phase-matching condition that relies on both the resonance condition of the ring and the interference pattern of the MZI. If the spectral width between two wavelengths of constructive interference in the MZI is twice the FSR of the ring, it is possible to suppress every second resonance of the ring. A schematic of such a device is shown in Figure \ref{fig:dmzr_schematic}a.

\begin{figure}[t]
\centering
\includegraphics[width=8cm]{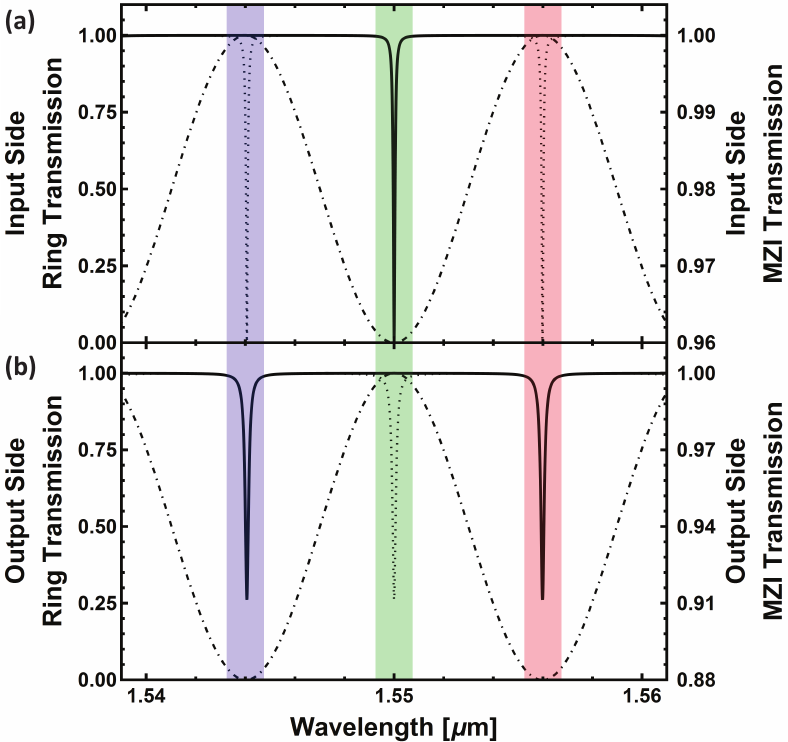}
\caption{Theoretical spectra of a microring resonator (dotted), Mach-Zehnder interferometer (dot-dashed), and a combination of the two (solid) for both (a) input and (b) output sides. The green, blue, and red shaded regions indicate the location of the pump, signal, and idler resonances respectively.}
\label{fig:dmzr_theory}
\end{figure}

For the case of a photon-pair source, one side of the ring can be used as the input for the pump photons and the drop side as the output for the generated photon-pairs. The MZI (MZI1) on the input side can be tuned to suppress every other resonance, while MZI2 on the output of the ring can be tuned to suppress the resonances allowed by MZI1 (i.e. they are perfectly out of phase with each other). This configuration will ensure the pump laser is critically coupled into the ring while not allowing it to exit out the drop port, and ensures that any photons that are generated at the resonances allowed by the drop port will only exit the over-coupled drop port (because MZI1 is tuned to not be phased matched with those photons). This makes the device function as though it is two independent single bus ring resonators, one for the input side and one for the output side. The input side ring is characterized by the transmission from the input port to the through port while the output side ring is characterized by the transmission from the add port to the drop port. The theoretical spectral response for both the input and output sides are shown in Figure \ref{fig:dmzr_theory}. This configuration has three key features: (i) The pump is critically coupled so the photon generation rate will be maximized; (ii) The pump is filtered from the photons that exit the drop port minimizing noise and reducing the amount of off-chip filtering required; (iii) The photon pairs will always leave out the same over-coupled drop port, yielding $\sim 100\%$ coincidence ratio, maximizing heralding efficiency. 

\begin{figure*}
\centering
\includegraphics[width=18cm]{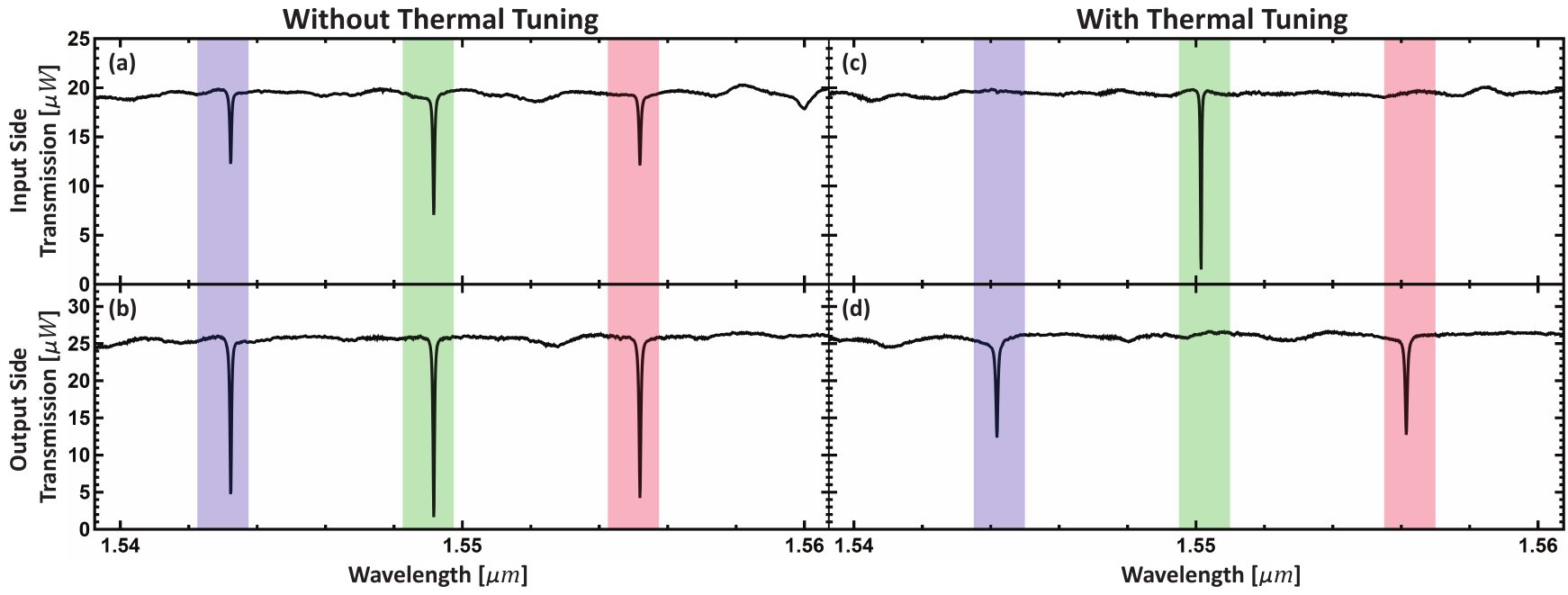}
\caption{Transmission spectra for the (a) input side and (b) output side of the DMZR without any thermal tuning. Transmission spectra for the (c) input side and (d) output side of the DMZR after optimization of the heaters. The green, blue, and red shaded regions indicate the location of the pump, signal, and idler resonances respectively.}
\label{fig:dmzr_spectrum}
\end{figure*}

\subsection{Dual Mach-Zehnder Microring Experimental Results}
\label{sec:DMZR_Exp}

\begin{table}[h]
\caption{Device Design Dimensions}
\label{table:dmzr_dimensions}
\begin{center}
\begin{tabular}{ |r|l| } 
 \hline
 Waveguide Width & $500 \nm$ \\
 \hline 
 Waveguide Thickness & $220 \nm$ \\
 \hline
 Ring Radius & $15 \um$ \\
 \hline
 Input Side Gap & $250\nm$ \\
 \hline
 Input Side Path Length Difference & $47.8 \um$ \\
 \hline
 Output Side Gap & $175 \nm$ \\
 \hline
 Output Side Path Length Difference & $48.0 \um$ \\
 \hline
\end{tabular}
\end{center}
\end{table}

An image of a fabricated device is shown in Figure \ref{fig:dmzr_schematic}b. The design dimensions for the device are shown in Table \ref{table:dmzr_dimensions}. Initial transmission spectra from each side of the device without any thermal tuning are shown in Figure \ref{fig:dmzr_spectrum}(a,b). In this configuration, all of the resonances are supported by both couplers. Upon performing a full 3-D sweep of voltages ranging from $0 \V$ to $10 \V$ across the three heaters, an optimum heater configuration (Ring Heater = $1 \V$, MZI1 Heater = $9 \V$, MZI2 Heater = $7 \V$) was found that exhibited the desired spectral dependence [Fig. \ref{fig:dmzr_spectrum}(c,d)].

\begin{figure}[b]
\centering
\includegraphics[width=8cm]{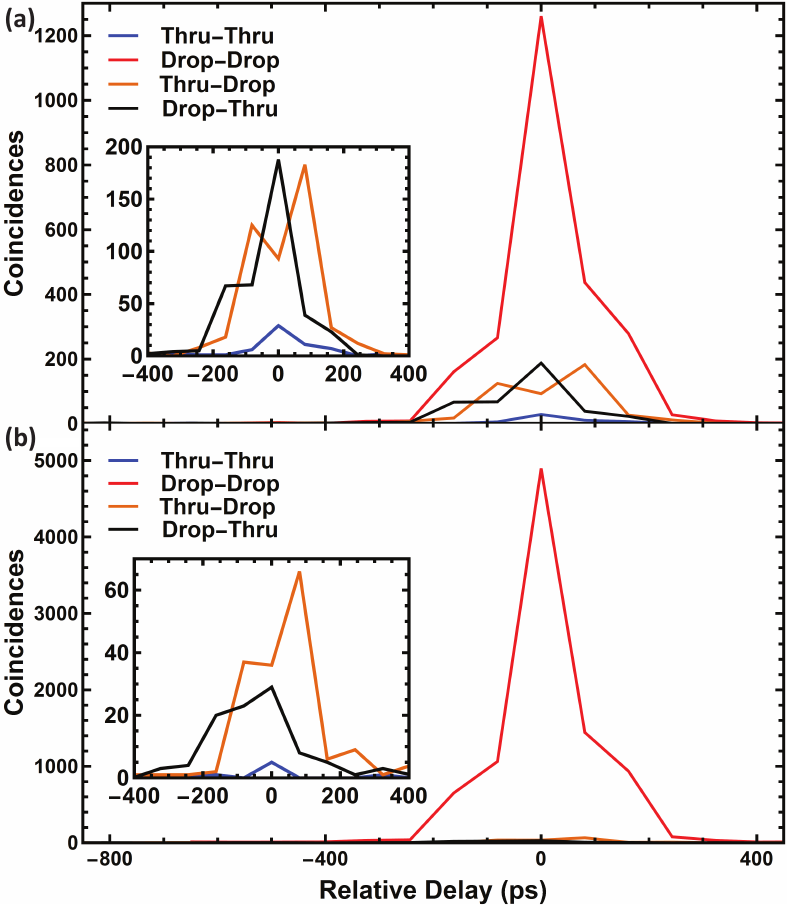}
\caption{Measured results from the dual Mach-Zehnder device showing the increase in coincidence counts when the resonances are (a) out of tune, (b) and tuned.}
\label{fig:dmzr_coincidences}
\end{figure}

A test setup similar to that shown in Figure \ref{fig:setup}(c) was used for time correlation measurements. Coincidences were collected for $60 \s$ with a timing resolution of $81 \ps$. The coincidence plots for the case where the heaters were not tuned are shown in Figure \ref{fig:dmzr_coincidences}(a). Due to the difference in the coupling gaps between the input side and the output side, the device already had an enhanced coincidence ratio of $\eta_\text{coinc}=0.73$ but the pump was not critically coupled to the ring [evident by the $\sim 4 \dB$ extinction in Figure \ref{fig:dmzr_spectrum}(a)]. The coincidence ratio was further improved in this configuration, resulting in a value of $0.97$. Even more notable is the enhancement to the pumping efficiency when compared to the untuned configuration. This is evident from both the $\sim 11 \dB$ extinction of the through port transmission and $3.7\, x$ increase in the drop port coincidence counts [shown in Fig. \ref{fig:dmzr_spectrum}(c) and Fig. \ref{fig:dmzr_coincidences}(b) respectively].

\section{Conclusion}
\label{sec:Conc}

In conclusion, this research shows that the coincidence ratio of a silicon ring resonator photon-pair source can be dramatically increased by engineering the coupling of a ring resonator. Two approaches were taken, the first decreased the input coupling to the ring but the pump power had to be correspondingly increased. However, this trade-off can be completely overcome with the addition of Mach-Zehnder legs attached to the resonator allowing for critical coupling of the pump into the resonator, and over coupling of the signal/idler photons through the drop port. This design enables photon-pair sources with inferred near maximal heralding efficiency out of the ring thanks to a measured coincidence ratio of $97\%$. The heralding efficiency is a very important characteristic of a photon-pair source, especially when studying the quantum mechanical properties of the photons. This is an essential and necessary step toward high performance and scalable quantum computers and communication systems.

\acknowledgements
This work was performed in part at the Cornell NanoScale Facility, a member of the National Nanotechnology Coordinated Infrastructure (NNCI), which is supported by the National Science Foundation (Grant ECCS-1542081). We acknowledge support for this work from the Air Force Research Lab (AFRL). This material is based upon work partially supported by the National Science Foundation under Award No. ECCS‐14052481. Any opinions, findings and conclusions or recommendations expressed in this material are those of the author(s) and do not necessarily reflect the views of AFRL or the National Science Foundation.


\end{document}